\documentclass[aps,pra,twocolumn,floatfix,showpacs,amsmath,amssymb]{revtex4}
\usepackage{graphicx}
\usepackage[hypertex]{hyperref}

\begin{document}

\newlength\smallfigwidth
\smallfigwidth=3.2 in

\preprint{Submitted to PRA}

\title{Resonant modes in triangular dielectric cavities}
\author{G.\ M.\  Wysin}
\email{wysin@phys.ksu.edu}
\homepage{http://www.phys.ksu.edu/~wysin}
\affiliation{
Department of Physics \\
Kansas State University \\
Manhattan, KS 66506-2601
}
\date{August 12, 2004}

\begin{abstract} 
The resonant optical modes of a high permittivity dielectric prism with
an equilateral triangular cross section are discussed.
Eigenmode solutions of the scalar Helmholtz equation with Dirichlet boundary 
conditions, appropriate to a conducting boundary, are applied for this purpose.
The particular plane wave components present in these modes are analyzed for their 
total internal reflection behavior and implied mode confinement when the conducting 
boundary is replaced by a sharp dielectric mismatch.
Improvement in TIR confinement by adjusting the longitudinal wavevector $k_z$ is
also discussed.
For two-dimensional electromagnetic solutions ($k_z=0$), TE polarization leads to 
longer lifetime than TM polarization, assuming that escape of evanescent boundary 
waves at the corners is the primary decay process.
\end{abstract} 

\maketitle

\section{Introduction}
Micro-sized dielectric cavities are becoming increasingly important, due to 
their applications especially in microlasers and resonators with various 
geometries, including disks,\cite{McCall92} triangles,\cite{Chang00}
squares,\cite{Poon01,Fong03,Moon03} and hexagons.\cite{Vietze98,Braun00}
Materials where the two-dimensional cross-section controls the properties include, 
for example, various cleaved semiconductor structures\cite{Chang00} as well as zeolite 
ALPO$_4$-5 crystals.\cite{Vietze98}
Even more interesting are lasing semiconductor pyramids that grow by natural
processes.\cite{pyramid1,pyramid2,pyramid3}

An understanding of the geometry dependence of the total internal reflection (TIR)
that produces resonant modes can lead to improvements in designs of optical devices.
For the reduced two-dimensional electrodynamics (fields independent of a
$z$-coordinate along an axis of symmetry), the boundary element 
method\cite{WiersigBEM} is very powerful for determining the scattering fields 
in TM or TE polarizations.
Alternatively, here we present a simpler analysis of the resonant modes based only 
on an analysis of the conditions needed for TIR within a dielectric cavity.
The calculation is approximate but simple, especially for a geometry where
the related Helmholtz equation has an exact analytic solution.

Here we consider the modes in equilateral triangular-based prisms and 
their two-dimensional (2D) analogs; the triangular system is chosen here for
its interesting symmetries and known analytic Helmholtz 
solution.\cite{Lame52,Krishna82,Itzyk86,BB97,Brack97,Sales03}
As a first approximation, a 2D dielectric system with Dirichlet boundary conditions 
(DBC) is considered, taking the fields equal to zero at the boundary, equivalent 
to a metallic or conducting boundary.
The exact analytic solutions for this 2D triangular system are reviewed; each mode
is a superposition of only six plane-wave components.
We describe how the mode information is useful for determining
which modes will be confined if the dielectric is surrounded
by a lower index dielectric (or vacuum), rather than a perfect conductor,
using the conditions for TIR.
A careful analysis shows that when a mode is strongly confined in the cavity
by TIR, Dirichlet boundary conditions apply (approximately) to the relevant Helmholtz
equation for both the TM and TE polarizations.
The simple plane wave components of each mode can be analyzed in terms of 
Maxwell's equations and the related Fresnel amplitude ratios, which are
slightly different for TM and TE polarizations.\cite{Jackson1} 
These differences are shown to imply longer estimated lifetimes for the TE modes, 
enhanced by a factor of the order of the squared index ratio 
$(\mathsf{n/n'})^2$, when compared to the lifetime of the corresponding 
TM mode.

Obviously some error is made compared to employing a more technically complete 
solution of Maxwell's equations such as boundary element methods;\cite{WiersigBEM} 
the fields discussed here for the dielectric mismatch boundary are not close to the 
true fields unless the mismatch is very large.
This approach, however, should approximately determine the symmetries of
the confined modes and give reasonable estimates for the dielectric mismatch 
needed for their confinements.
The improvement of confinement for finite--height prisms with the same cross 
sections, allowing for nonzero longitudinal wavevector $k_z$,  is also discussed.

\section{Simplified Quasi-Two-Dimensional Helmholtz Problems}
Within a cavity with electric permittivity $\epsilon$ and 
magnetic permeability $\mu$, we consider the solutions of a scalar 
wave equation for any component $\psi$ of the electric or magnetic field,
\begin{equation}
\label{wave}
\nabla^2 \psi 
- \frac{\epsilon\mu}{c^2} \frac{\partial^2 }{\partial t^2} \psi
= 0,
\end{equation}
where $c$ is the speed of light in vacuum and $t$ is time.
The cavity shape being considered is a prism with an equilateral triangular 
cross-section of edge $a$ in the $xy$-plane; the height is $h$ along the $z$ axis.
Ultimately, we want to discuss which modes will be confined in the cavity of 
index of refraction $\mathsf{n}=\sqrt{\epsilon\mu}$ when it is surrounded by 
a uniform medium of lower index of refraction 
$\mathsf{n^{\prime}}=\sqrt{\epsilon^{\prime}\mu^{\prime}}$ (for example, vacuum).

We first analyze the modes, assuming the fields go to zero at the cavity walls, 
using Dirichlet boundary conditions (DBC), which corresponds to perfectly
conducting walls, or a mirrored cavity. 
The range of applicability of these boundary conditions for a mode that is 
confined by TIR caused by an index mismatch, rather than reflecting boundaries,
is discussed subsequently.
A solution at frequency $\Omega$ is sought, with $e^{-i \Omega t}$ time dependence, 
corresponding to three-dimensional (3D) wavevectors with magnitude
\begin{equation}
\label{kwave}
K = \frac{\Omega}{c^*}, \quad\quad c^{*}=\frac{c}{\sqrt{\epsilon\mu}};
\end{equation}
the speed of light in the cavity medium is denoted as $c^*$.
Then we are solving an eigenvalue problem (Helmholtz equation)
within the given geometry,
\begin{equation}
\label{Helmholtz}
\nabla^2 \psi = - K^2 \psi.
\end{equation}
In the general case, we can consider that we are looking for a
superposition of plane waves in the form 
$e^{i (\vec{K}\cdot\vec{r} -\Omega t)}$, all of which  
satisfy \eqref{Helmholtz}, and with the correct linear combination
to satisfy the required boundary conditions.
These different components have the same squared wavevector, although they 
can possess different directions of propagation and different amplitudes.
The symmetry of the situation, however, requires only some reduced set of 
possible wavevectors for triangular-based prisms.

A wavevector is composed from a $z$-component, $k_z$,  along the vertical or 
longitudinal axis, and the remaining 2D components in the $xy$ plane,
$k^2 \equiv k_x^2 + k_y^2$, so that we decompose the squared magnitude as
\begin{equation}
K^2 = k_z^2 + k^2.
\end{equation}
Then the frequency is partitioned accordingly,
\begin{equation}
\Omega^2 = \omega_z^2 + \omega^2,
\end{equation}
where the $xy$ frequency $\omega$ and wavevector $k$ are
related by an equation analogous to \eqref{kwave}, as $k=\omega/c^*$.
The remaining part, $\omega_z$, is determined for a prism of height $h$, 
assuming some $\psi\propto sin(k_z z)$ longitudinal dependence, by 
\begin{equation}
k_z = \frac{l\pi}{h},  \quad\quad l = 1, 2, 3 ...
\end{equation}
which automatically enforces DBC at the ends of the prism.
As a result, for the prism geometry, we only solve the separated 
Helmholtz equation for two dimensions,
\begin{equation}
\label{2DHelmholtz}
\nabla_{xy}^2 \psi = - k^2 \psi.
\end{equation}
%

\subsection{2D Electromagnetics at a boundary between media}
\label{2DEM}
For 2D E\&M problems, which are defined by no dependence of
the fields along the prism axis, one takes $k_z=0$, ignoring any
boundary conditions at the prism ends (which can be considered
at infinity). 
Then Maxwell's equations and their associated physical boundary conditions
show that one need only consider transverse magnetic (TM) and transverse 
electric (TE) polarizations of the fields.  
(Presence of nonzero $k_z$ for a dielectric waveguide surrounded by
a different dielectric medium, in general, does not lead to separated TM
and TE modes, see Ref. \onlinecite{Jackson1}.)
It is well-known that in a \emph{conducting} waveguide or resonator, at the 
boundaries of the 2D cross-section, the field $\psi=E_z$ for TM modes satisfies 
DBC, and the field $\psi=B_z$ for TE modes satisfies Neumann boundary conditions 
(NBC).
Alternatively, if the medium is surrounded simply by a different 
(nonconducting) dielectric medium,  there still are TM or TE polarizations,
but, strictly speaking, these would satisfy the more involved boundary conditions
of Maxwell's equations, rather than the oversimplified DBC or NBC.  
If the fields are strongly undergoing TIR (incident angle well beyond the 
critical angle), however, both polarizations are shown here to satisfy DBC, 
to a certain limited extent.  
This can be seen by examining the actual vector fields present under the reflection 
and refraction at a boundary between two media, for the two possible polarizations.

Consider a planar boundary between two media, where the boundary defines the 
$xz$-plane.
The region $y<0$ is occupied by a medium of index $\mathsf{n}=\sqrt{\epsilon\mu}$, 
while the region $y>0$ is occupied by a medium of index 
$\mathsf{n}'=\sqrt{\epsilon'\mu'}$, with $\mathsf{n}>\mathsf{n}'$.
Primes refer to quantities on the refracted wave side (ultimately, \emph{outside} 
the system we are studying).
The wavevector magnitudes are $k=\frac{\omega}{c}\sqrt{\epsilon\mu}$ and
$k'=\frac{\omega}{c}\sqrt{\epsilon'\mu'}$ in the two media.
All the waves in this discussion are assumed to have $e^{-i\omega t}$ time 
dependence.
A 2D plane wave $\sim e^{i(k_x x + k_y y)}$ with fields $\vec{E}_i, \vec{B}_i$, 
propagating in medium $\mathsf{n}$ with 
$\vec{k}_i=(k_x,k_y)=k(\sin\theta_i, \cos\theta_i)$ and incident on the 
boundary, can be polarized in two ways, which correspond exactly to the TM and 
TE polarizations.

Diagrams of the two possible situations are given, for example, in Chap.\ 7 of 
Ref.\ \onlinecite{Jackson1}.
For $\vec{E}_i$ perpendicular to the plane of incidence ($xy$-plane), $\vec{E}_i=E_i\hat{z}$,
and the magnetic field will have no $z$-component anywhere, which is the same as the 
TM polarization.
On the other hand, for $\vec{E}_i$ lying within the plane of incidence, 
$\vec{B}_i=B_i \hat{z}$, and the electric field has no $z$-component anywhere, 
corresponding to the TE polarization.  
Thus, we can use the  Fresnel reflection and refraction amplitudes for these 
two cases, that result from the actual boundary conditions for Maxwell's 
equations, to investigate the appropriateness of applying DBC for fields 
undergoing TIR.
Furthermore, this analysis will be used for determining the properties of the
evanescent refracted wave under TIR;  the power lost in this boundary wave can
be expected to play a role in the lifetime of a resonant mode.

\vskip 0.1in \noindent
\textbf{TM polarization:}  
For incident $\vec{E}_i=E_i^{0} \hat{z} e^{i(k_x x + k_y y)}$ 
perpendicular to the plane of incidence, there is also a reflected wave 
$\vec{E}_r = E_r^{0} \hat{z} e^{i(k_x x -k_y y)}$ with the same polarization, but a 
different magnitude and phase, depending on the angle of incidence $\theta_i$.  
One can express the reflected wave amplitude $E_r^{0}$ by the Fresnel formula:
\begin{subequations}
\begin{equation}
E_r^{0}=E_i^{0} ~ e^{-i\alpha},
\end{equation}
\begin{equation}
e^{-i \alpha} = \frac{\sqrt{\frac{\epsilon}{\mu}}\cos\theta_i 
                     -\sqrt{\frac{\epsilon'}{\mu'}}\cos\theta'}
                     {\sqrt{\frac{\epsilon}{\mu}}\cos\theta_i 
                     +\sqrt{\frac{\epsilon'}{\mu'}}\cos\theta'}
\end{equation}
\end{subequations}
Here $\theta'$ is the angle of the refracted wave, obtained from Snell's Law,
$n\sin\theta_i = n' \sin\theta'$,  which implies TIR when $\theta_i$ surpasses
the critical angle $\theta_c$, defined by
\begin{equation}
\sin\theta_c=\mathsf{\frac{n'}{n}}.
\end{equation}

Under TIR, $E_r^{0}$ has the same magnitude as $E_i^{0}$, but is phase shifted 
by angle $\alpha$, because the cosine of the refracted wave becomes pure imaginary:
\begin{equation}
\cos\theta' = i\gamma, \quad \gamma=\sqrt{(\sin\theta_i/\sin\theta_c)^2-1}.
\end{equation}
Then it is useful to express the phase difference between the incident and
reflected waves as
\begin{equation}
\label{alphaTM}
\tan\frac{\alpha}{2} = \frac{\mu}{\mu'} \sqrt{\frac{\cos^{2}\theta_c}{\cos^{2}\theta_i}-1}
\end{equation}
The linear combination of incident and reflected waves in medium $\mathsf{n}$ is
$\vec{E}=\vec{E}_i+\vec{E}_r = E_z \hat{z}$, having the spatial variation
approaching the boundary (region $y<0$),
\begin{equation}
\label{EzTM}
E_z=E_i+E_r= 2 E_i^{0} e^{-i\frac{\alpha}{2}} e^{ik_x x} 
                       \cos\left(k_y y +\frac{\alpha}{2}\right)
\end{equation}
Corresponding to this is the associated evanescent wave in medium $\mathsf{n}'$ 
(region $y>0$), 
\label{E'TM}
\begin{equation}
E_z' = 2 E_i^{0} e^{-i\frac{\alpha}{2}} e^{ik_x x} e^{-k' \gamma y},
\end{equation}
where $k_x=k_x'$ due to Snell's Law.
Inspection of Eq.\ \eqref{EzTM} shows that the $E_z$ field on the incident side 
acquires a node at the boundary $y=0$ and satisfies Dirichlet BC only when the 
phase shift attains the value $\alpha=\pi$.  
According to \eqref{alphaTM}, this occurs only in the limit $\theta_i\to 90^{\circ}$,
i.e., extreme grazing incidence.
Alternatively, at the \emph{threshold} for TIR ($\theta=\theta_c$), \eqref{alphaTM} 
gives $\alpha=0$, whereby \eqref{EzTM} indicates that the $E_z$ field now will peak 
at $y=0$ and satisfies a Neumann BC.  
%

Seeing that the BC on $E_z$ ranges between the two extremes of DBC and NBC, clearly
neither boundary condition is fully applicable to the TIR regime.
However, it suggests that one should use NBC for finding the TIR threshold conditions,
and DBC to determine the field distributions at large enough index mismatch.
One can get some indication of the crossover to DBC-like behavior by locating 
the incident angle $\theta_i=\tilde{\theta}$ where the phase shift passes 
$\frac{\pi}{2}$, i.e., when $\tan\frac{\alpha}{2} = 1$.
From \eqref{alphaTM} one finds 
\begin{equation}
\cos\tilde{\theta} = \frac{\cos\theta_c}{\sqrt{1+\left(\frac{\mu'}{\mu}\right)^2}}.
\end{equation}
In the usual practical situation, with $\mu=\mu' \approx 1$, we have 
$\cos\tilde{\theta}=\frac{1}{\sqrt{2}}\cos\theta_c$. 
For usual optical media with very weak magnetic properties, one sees that a large
index ratio $\mathsf{n/n'}\approx \sqrt{\epsilon/\epsilon'}$ does not increase 
the range of applicability of Dirichlet BC for TM polarization.
To give a numerical example, for a weak index mismatch with $\sin\theta_c=1/2$, 
giving $\theta_c=30^{\circ}$, the crossover angle is $\tilde{\theta}=52.2^{\circ}$;
in order to reach phase angle $\alpha =0.9 \pi$, quite close to DBC, requires an
incident angle $\theta_i=82.2^{\circ}$. 
Even for a larger mismatch $\sin\theta_c=1/4$, with $\theta_c=14.5^{\circ}$, one
gets only a slight improvement to $\tilde{\theta}=46.8^{\circ}$, and to get to
$\alpha=0.9 \pi$ still requires $\theta_i=81.3^{\circ}$.

The conclusion is that it makes some reasonable sense to apply NBC to find the
limiting conditions for TIR, but, in general, provided the incident angle is 
sufficiently larger than the crossover angle $\tilde{\theta}$, an approximate 
description of the fields on the incident side should be possible using DBC.
The application of DBC to this problem improves with higher index mismatch,
but not as strongly as one would hope, because the phase angle does not
depend on the dielectric permittivities for this polarization.

\vskip 0.1in \noindent
\textbf{TE polarization:} 
The discussion is similar, but now the magnetic field $\vec{B}$ is 
polarized in the $\hat{z}$ direction everywhere, and controls the other fields, 
according to relations $\vec{B}=\sqrt{\epsilon\mu}~ \hat{k}\times\vec{E}$ for 
each plane wave, and amplitude relations $B=\sqrt{\epsilon\mu}~E$.  
Taking incident wave $\vec{B}_i=B_i^{0} \hat{z} e^{i(k_x x + k_y y)}$, with
electric field amplitude $E_i^{0}=B_i^{0}/\sqrt{\epsilon\mu}$, 
there is a reflected wave $\vec{B}_r = B_r^{0} \hat{z} e^{i(k_x x -k_y y)}$,
with electric field amplitude $E_r^{0}=B_r^{0}/\sqrt{\epsilon\mu}$.
Now the amplitude ratio $E_r^{0}/E_i^{0}=e^{-i\alpha}$ is described by the 
different Fresnel formula:
\begin{equation}
e^{-i \alpha} = \frac{\sqrt{\frac{\epsilon'}{\mu'}}\cos\theta_i 
                     -\sqrt{\frac{\epsilon}{\mu}}\cos\theta'}
                     {\sqrt{\frac{\epsilon'}{\mu'}}\cos\theta_i 
                     +\sqrt{\frac{\epsilon}{\mu}}\cos\theta'}
\end{equation}
Clearly, the same amplitude ratio also applies to $B_r^{0}/B_i^{0}$.
The phase difference between the incident and reflected waves can be expressed as
\begin{equation}
\label{alphaTE}
\tan\frac{\alpha}{2} = \frac{\epsilon}{\epsilon'} 
                       \sqrt{\frac{\cos^{2}\theta_c}{\cos^{2}\theta_i}-1}.
\end{equation}
The linear combination of incident and reflected magnetic waves in medium 
$\mathsf{n}$ is $\vec{B}=\vec{B}_i+\vec{B}_r = B_z \hat{z}$, and mirrors the 
behavior of $\vec{E}$ for the TM problem, having the spatial variation approaching 
the boundary,
\begin{equation}
\label{BzTE}
B_z=B_i+B_r= 2 B_i^{0} e^{-i\frac{\alpha}{2}} e^{ik_x x} 
                       \cos\left(k_y y +\frac{\alpha}{2}\right)
\end{equation}
The associated evanescent wave in medium $\mathsf{n}'$ is
\label{B'TE}
\begin{equation}
B_z' = 2 B_i^{0} e^{-i\frac{\alpha}{2}} e^{ik_x x} e^{-k' \gamma y}.
\end{equation}
Clearly, the behavior of $B_z$ near the boundary for TE polarization is the same 
as that for $E_z$ near the boundary for TM polarization.  
It means that provided the incident angle is far enough beyond the critical angle, 
one should also apply DBC for TE fields; NBC would only be reasonable just beyond 
the TIR threshold.
However, due to the presence of the permittivity ratio in \eqref{alphaTE}, a large 
index mismatch \emph{does} enhance the applicability of DBC for TE polarization.
The crossover incident angle $\tilde{\theta}$ at which $\tan\frac{\alpha}{2}=1$ 
is now
\begin{equation}
\cos\tilde{\theta} = \frac{\cos\theta_c}{\sqrt{1+\left(\frac{\epsilon'}{\epsilon}\right)^2}}
  \approx \frac{\cos\theta_c}{\sqrt{1+\left(\mathsf{\frac{n'}{n}}\right)^4}},
\end{equation}
where the latter expression applies when $\mu \approx \mu'$.  
For some numerical examples, now the modest index mismatch with $\sin\theta_c=1/2$ 
and $\theta_c=30^{\circ}$ leads to a very nearby crossover angle 
$\tilde{\theta}=32.8^{\circ}$, meaning that the strong TIR regime and region of 
adequate applicability of DBC is very wide.
For larger mismatch $\sin\theta_c=1/4$ with $\theta_c=14.5^{\circ}$, the DBC 
regime is even closer to $\theta_c$, beginning around $\tilde{\theta}=14.9^{\circ}$.  
Thus, application of Dirichlet BC to finding the 2D resonant modes for a cavity with
TE field polarization should be very acceptable, even more so than for TM 
polarization, except when the mode has plane wave components extremely close to 
their TIR threshold.

\vskip 0.1in
Within the limitations indicated in the above analysis, we continue by discussing
the modes in a 2D equilateral triangle, under the assumption of DBC for either
the TM or TE polarizations.

\subsection{Exact modes for an equilateral triangle with DBC}
\label{Tri2D}

\begin{figure}
\includegraphics[angle=0.0,width=\columnwidth]{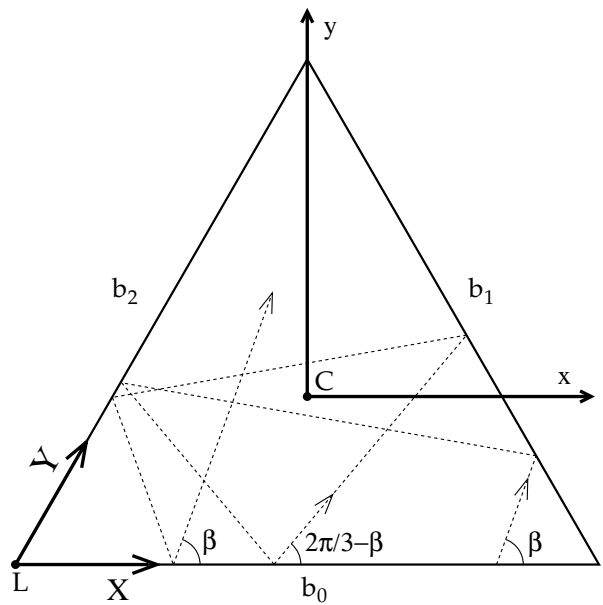}
\caption{
\label{trisys}
Description of the coordinates for a 2D triangular cavity of edge $a$. The 
geometrical center at $C$, the origin of the $xy$-coordinates, is a distance
$\frac{a}{2\sqrt{3}}$ above the lower ($b_0$) edge.  The lower left corner at $L$ is
the origin of the skew $XY$-coordinates.  The dashed line demonstrates the 
reflections of a ray originating at angle $\beta=70^{\circ}$ to the lower edge,
requiring two complete circuits to return to the same angle. 
}
\end{figure}

The triangular cross section offers an opportunity for exact solutions
to the 2D Helmholtz equation.
This problem has been solved analytically\cite{Lame52,Krishna82,Itzyk86,BB97} for 
both DBC and NBC in several contexts, including quantum billiards 
problems,\cite{Gutkin86,Liboff94,Molinari97,Sales03}
quantum dots,\cite{Brack97} and lasing modes in 
resonators and mirrored dielectric cavities.\cite{Frank96,Chang00} 
In particular, it is interesting to note that only the triangles with
angle sets $\pi(\frac{1}{3},\frac{1}{3},\frac{1}{3})$, 
$\pi(\frac{1}{2},\frac{1}{4},\frac{1}{4})$,
and $\pi(\frac{1}{2},\frac{1}{3},\frac{1}{6})$ are classically 
integrable\cite{Liboff94} and have simple closed form wavefunction solutions
derived in various ways,\cite{Krishna82,Itzyk86,Molinari97} following
the first solution for triangular elastic membranes by Lam\'e.\cite{Lame52}
Here we use the equilateral triangle of edge length $a$ for its interesting 
symmetries and resulting simplifications.

Coordinates are used where the origin is placed at the geometrical center of
the triangle, and the lower edge is parallel to the $x$-axis, as shown in
Fig.\ \ref{trisys}.
The notation $b_0$, $b_1$, and $b_2$ is used to denote
the lower, upper right, and upper left boundaries, respectively.
This discussion concerns DBC.

Following Chang \textit{et al.},\cite{Chang00} some comments can be made on the
sequence of reflections a plane wave trapped in the cavity makes, but with
a slightly different physical interpretation.
A plane wave leaving the lower face ($b_0$) at angle $\beta$ to the x-axis 
sequentially undergoes reflections at the other two faces, generating plane waves 
at angles $240^{\circ}-\beta$, $\beta-120^{\circ}$, relative to the x-axis.
Finally, when reflected again off the lower face, it comes
out at $120^{\circ}-\beta$, which does not match the original
wave unless $\beta=60^{\circ}$.
However, when allowed to propagate \emph{again} around the triangle,
the sequence of angles is $120^{\circ}+\beta$, $-\beta$, which then
comes out at $+\beta$ after the reflection off the lower face.
So the wave closes on itself after two full revolutions,
no matter what the initial angle. 
As an example, the sequence of reflections starting with $\beta=70^{\circ}$
from boundary $b_0$ are shown in Fig.\ \ref{trisys}.
In general, any original wave simply generates a set of six symmetry related 
waves rotated by $\pm 120^{\circ}$ and inversions through the y-axis.

\begin{figure}
\includegraphics[angle=-90.0,width=\columnwidth]{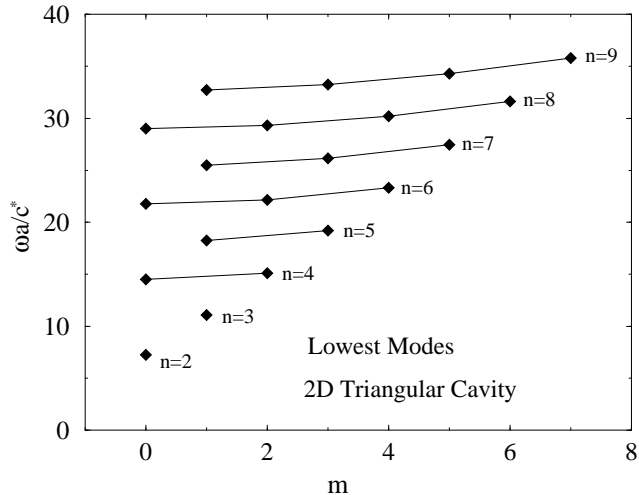}
\caption{
\label{trifreqs}
Scaled frequencies of some of the lowest possible modes in a 2D triangular 
system, as a function of quantum index $m$, for indicated $n>m$.  
Indices $n$ and $m$ must be integers of equal parity. $c^*$ is the light
speed in the medium.
}
\end{figure}

These considerations show that the general solution is a superposition of 
six plane waves, which can be obtained by 120-degree rotations
of one partially standing wave.
Consider initially a wavevector with $xy$ Cartesian components, 
$\vec{k}=(k_1,k_2)$, and defined basic vectors $\vec{k}_1 = k_1 \hat{x}$, 
$\vec{k}_2 = k_2 \hat{y}$, producing a wavefunction written as
\begin{equation}
\label{psi0}
\psi_0 = e^{i \vec{k}_1\cdot\vec{r}} 
         \sin\left[ \vec{k}_2\cdot\vec{r} + \frac{k_2 a}{2\sqrt{3}} \right]
\end{equation}
This is a combination of plane waves at two
angles $\beta$ and $-\beta$ to the x-axis, where 
$\tan\beta = k_2/k_1$, and the combination goes to zero 
on boundary $b_0$, where $y=-\frac{a}{2\sqrt{3}}$;
thus it is like a standing wave.  
By assumption, we need $k_2\ne 0$ to have a nonzero wavefunction.
As discussed above, when reflected off the triangle faces $b_1$ and $b_2$, 
this simply produces rotations of $\pm 120^{\circ}$.  
If $R$ is an operator that rotates vectors through $+120^{\circ}$,
the additional waves being generated are
\begin{equation}
\label{psi1}
\psi_1 = e^{i (R\vec{k}_1)\cdot\vec{r}} 
         \sin \left[ (R\vec{k}_2)\cdot\vec{r} + \frac{k_2 a}{2\sqrt{3}} \right]
\end{equation}
\begin{equation}
\label{psi2}
\psi_2 = e^{i (R^2\vec{k}_1)\cdot\vec{r}} 
         \sin \left[ (R^2\vec{k}_2)\cdot\vec{r} + \frac{k_2 a}{2\sqrt{3}} \right]
\end{equation}
where the rotated vectors are
\begin{eqnarray}
R\vec{k}_1 &=& k_1 \left( -\frac{1}{2}\hat{x}+\frac{\sqrt{3}}{2}\hat{y} \right) \nonumber \\
R^2\vec{k}_1 &=& k_1 \left( -\frac{1}{2}\hat{x}-\frac{\sqrt{3}}{2}\hat{y} \right)
\end{eqnarray}
\begin{eqnarray}
R\vec{k}_2 &=& k_2 \left( -\frac{\sqrt{3}}{2}\hat{x}-\frac{1}{2}\hat{y} \right) \nonumber \\
R^2\vec{k}_2 &=& k_2 \left( \frac{\sqrt{3}}{2}\hat{x}-\frac{1}{2}\hat{y} \right) 
\end{eqnarray}
By design, these rotated waves satisfy the boundary conditions, $\psi_1(b_1) = 0$
and $\psi_2(b_2) = 0$.

\begin{figure}
\includegraphics[angle=0.0,width=\smallfigwidth]{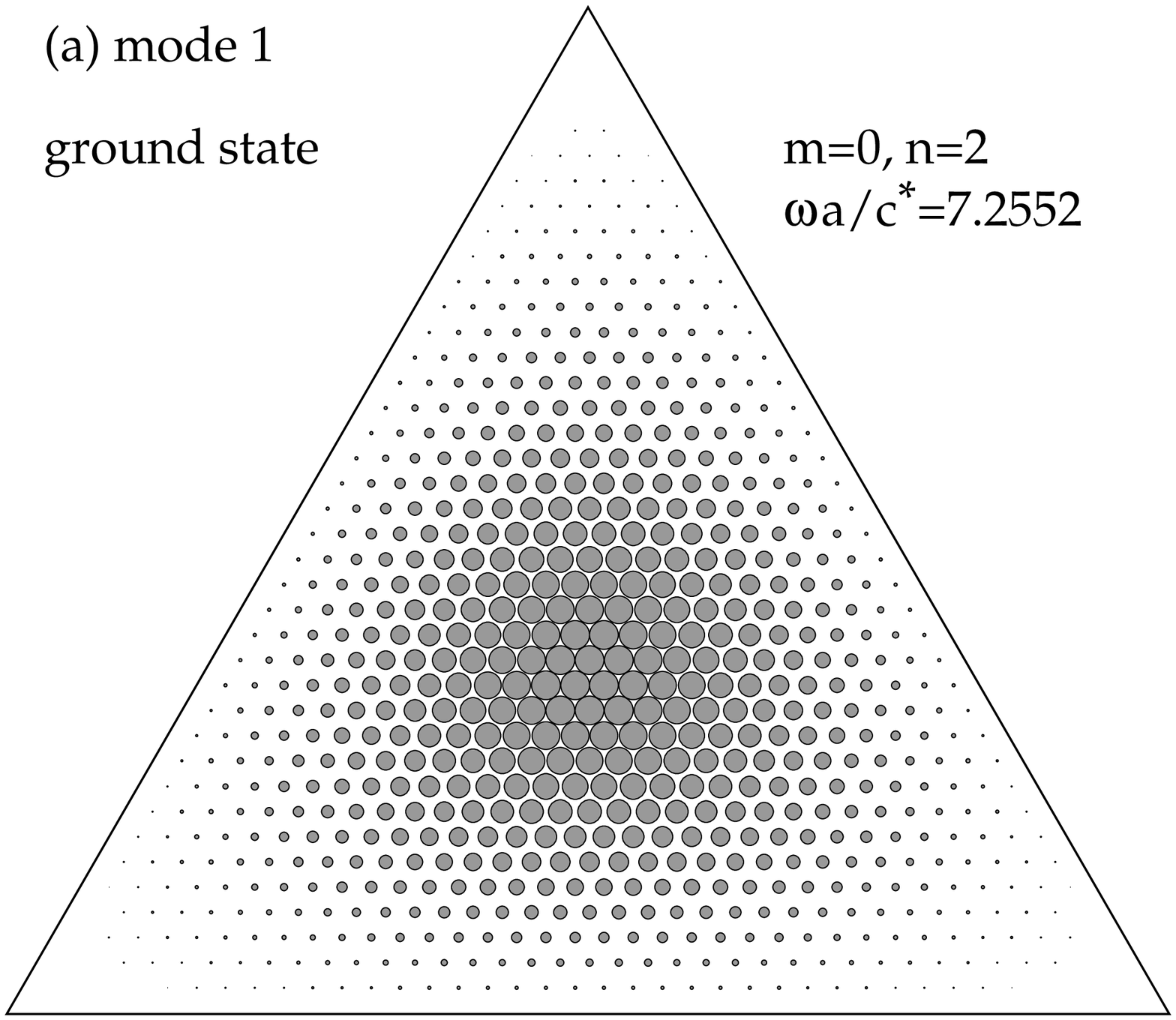}
\includegraphics[angle=0.0,width=\smallfigwidth]{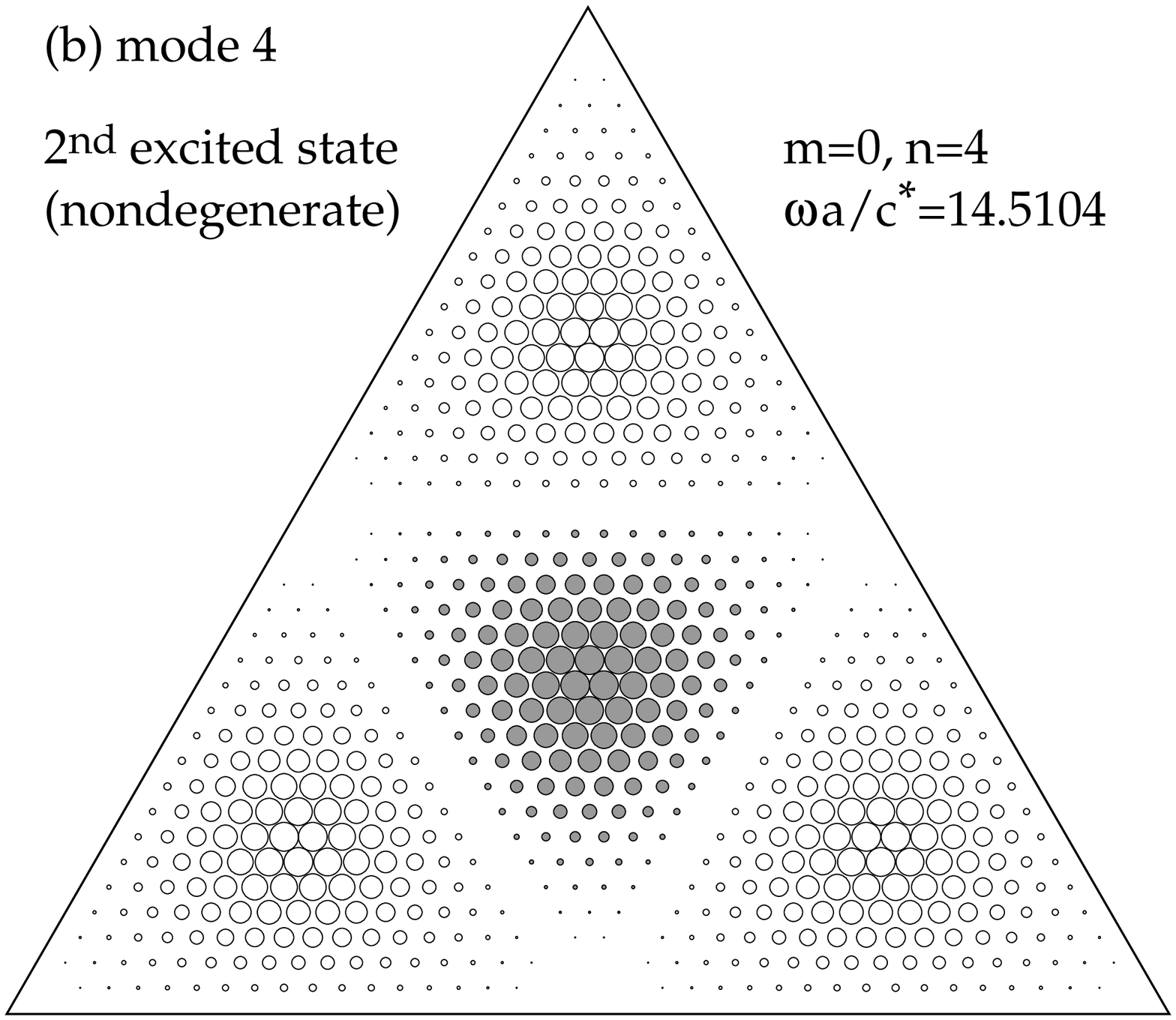}
\caption{
\label{trimodesA}
Wavefunctions of the two lowest modes with $m=0$ in a 2D triangular system:
the ground state (a), with $n=2$, and the third excited state (b), with $n=4$, 
at double the frequency of the ground state.  Solid/partial shadings represent 
positive/negative wavefunction values, whose magnitudes correspond to the radii 
of the symbols.  A grid with $N=40$ was only used to present the diagrams.  Grid 
sites without symbols have $|\psi|<0.02|\psi_{\textrm{max}}|$. 
}
\end{figure}

In order to determine the allowed $(k_1,k_2)$, one needs to 
impose DBC on all three boundaries, for a linear combination
with unknown coefficients ${\cal A}_0, {\cal A}_1, {\cal A}_2$,
\begin{equation}
\label{triwf}
\psi = {\cal A}_0\psi_0+{\cal A}_1\psi_1+{\cal A}_2\psi_2 .
\end{equation}
Imposing DBC on all boundaries determines the allowed wavevector components as
\begin{equation}
\label{k1values}
k_1  = \frac{2\pi}{3a} m, \quad \quad m=0,1,2...
\end{equation}
\begin{equation}
\label{k2values}
k_2 = \frac{2\pi}{3a} \sqrt{3}~n, \quad\quad n=1,2,3...
\end{equation}
Furthermore, the parity constraint $e^{i\pi m}=e^{i\pi n}$ appears; 
that is, $n$ and $m$ are either both odd or both even.

The resulting $xy$ frequencies are given by
\begin{equation}
\label{2domega}
\omega = c^{*} \sqrt{k_1^2+k_2^2}
       = \frac{c}{\sqrt{\epsilon\mu}} \frac{2\pi}{3a}\sqrt{m^2+3n^2}
\end{equation}
The mode wavefunctions are described completely using the amplitude
relationships that result:
\begin{equation}
{\cal A}_1 = {\cal A}_0 e^{i \frac{2\pi}{3} m}, \quad\quad
{\cal A}_2 = {\cal A}_0 e^{-i\frac{2\pi}{3} m}.
\end{equation}
Therefore, solutions are specified by a choice of integers 
$m$ and $n$ and the phase of the complex constant ${\cal A}_0$.  
In Fig.\ \ref{trifreqs} the frequencies of the lowest modes
are presented, scaled with the speed of light in the medium, $c^{*}$,
and arranged as families for each value of the quantum number $n$. 
This solution is represented in the physically motivated form described by 
Chang {\it et al.},\cite{Chang00} and is entirely equivalent to the first 
solution given by Lame'\cite{Lame52} and revisited by various 
authors.\cite{BB97}

\begin{figure}
\includegraphics[angle=0.0,width=\smallfigwidth]{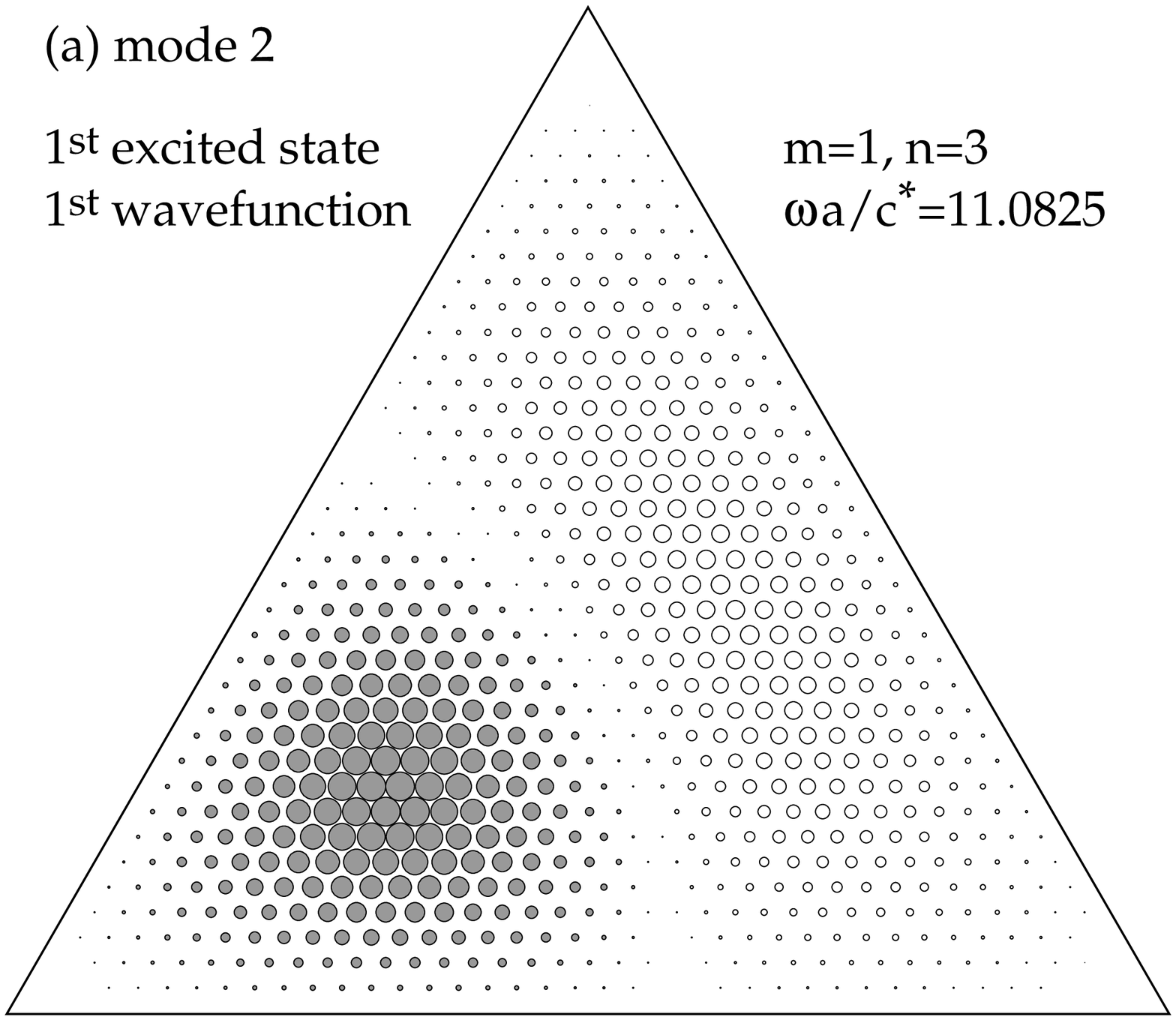}
\includegraphics[angle=0.0,width=\smallfigwidth]{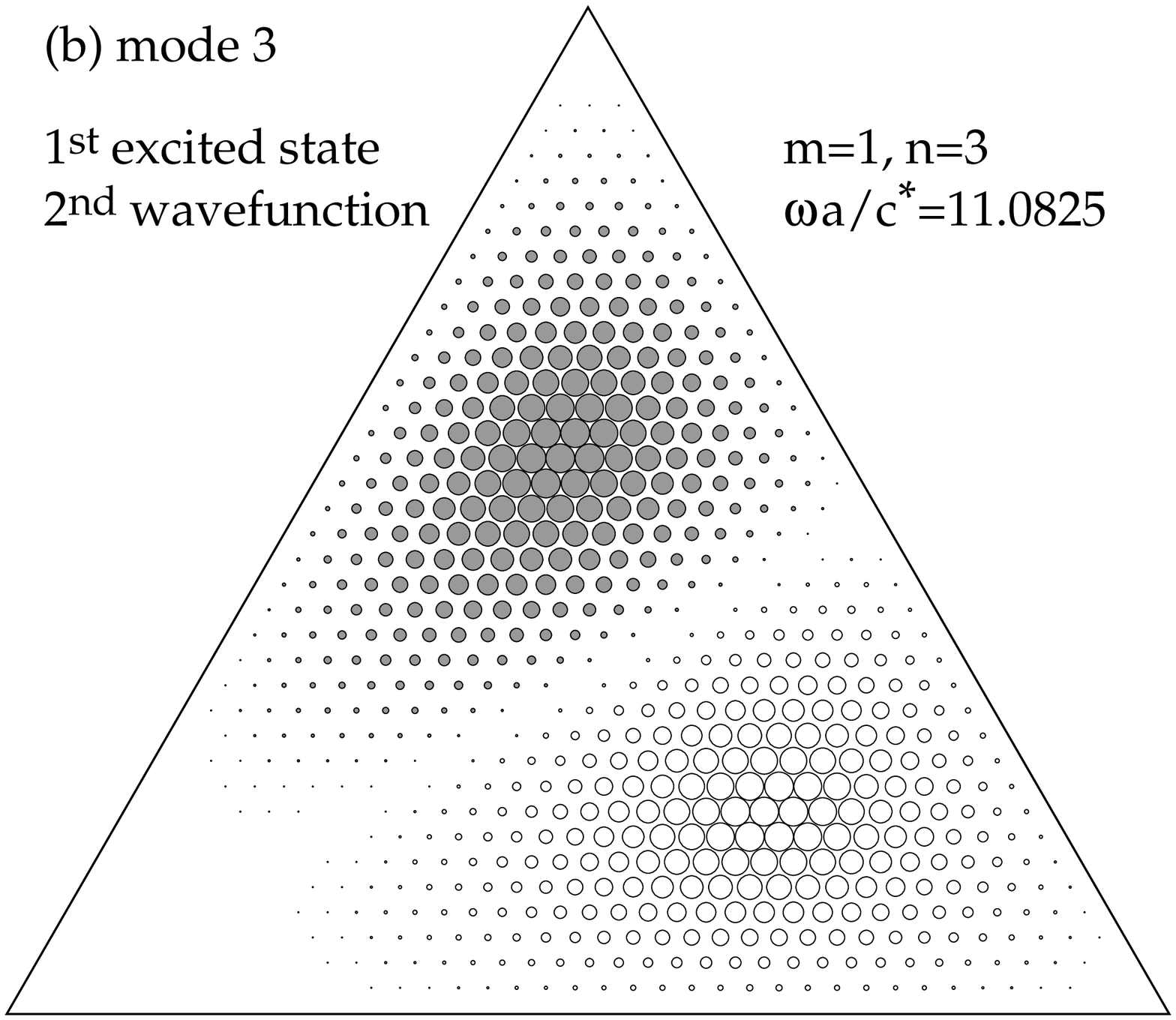}
\caption{
\label{trimodesB}
Wavefunctions of the doubly degenerate first excited state in a 2D 
triangular system.  Two orthogonal sub-states corresponding to 
different choices of the phase of $\psi$ are presented in (a) and (b).
}
\end{figure}

The solutions obtained have obvious symmetry properties.
We can get all the possible eigenfrequencies by applying the
restrictions, $0\le m<n$.  
Choices with $m\ge n$ or $m<0$ also give allowed frequencies, however, 
these only correspond to other $\vec{k}$ rotated by $\pm 120^{\circ}$
from some original $\vec{k}$ defined with $0\le m < n$.
For instance, having found $\vec{k}=(k_1,k_2)=\frac{2\pi}{3a}(m,\sqrt{3}n)$,
values of $\vec{k}^{\prime} = (k_1^{\prime},k_2^{\prime})
=\frac{2\pi}{3a} (m^{\prime},\sqrt{3} n^{\prime})$ corresponding to 
$\pm 120^{\circ}$ rotations result from
\begin{eqnarray}
k_1^{\prime} &=& k_1 \cos 120^{\circ} \mp k_2 \sin 120^{\circ} \nonumber \\
k_2^{\prime} &=& \pm k_1 \sin 120^{\circ} + k_2 \cos 120^{\circ}
\end{eqnarray}
which implies transformation to the new mode indexes,
\begin{eqnarray}
m^{\prime} = -\frac{1}{2} m \mp \frac{3}{2} n  \nonumber \\
n^{\prime} = \pm \frac{1}{2} m - \frac{1}{2} n .
\end{eqnarray}
As specific examples, the choice $(m,n)=(1,3)$ gives one mode,
which after rotations by $\pm 120^{\circ}$ could also be described
by the integers $(m,n)=(4,2)$ or by $(m,n)=(5,1)$.  
Note, however, this represents only one mode, with different choices
of the reference edge or triangle base.
We should also note, any mode with $m\ne0$ is doubly degenerate;
the change $m\rightarrow -m$ gives an independent mode with the
same frequency, which rotates in the opposite sense around the 
triangle.  
A real representation of the degenerate pairs comes from taking the
real and imaginary parts of $\psi$ to form two independent wavefunctions.
Different values of the complex constant ${\cal A}_0$ will produce other 
choices of the two independent wavefunctions.
For $m=0$, the wavefunction $\psi$ can be made pure real; the modes are 
non-degenerate and $n$ must be even, since $m$ and $n$ must be of equal parity.

The ground state (Fig.\ \ref{trimodesA}a) has $(m,n)=(0,2)$ and corresponding $xy$
frequency, $\omega = \frac{c}{\sqrt{\epsilon\mu}}\frac{4\pi}{\sqrt{3}a}$.
This corresponds to a wavelength of 
$\lambda = \frac{2\pi}{k} = \frac{\sqrt{3}a}{2}$, which is equal
to the height of the triangle.  
Naively, one might have expected that a half-wavelength of the
lowest mode could have fit across the triangle, as would occur 
in simple rectangular or one-dimensional geometry.  
Apparently, that choice would not have the ability to fully satisfy
DBC on all three surfaces.

\begin{figure}
\includegraphics[angle=0.0,width=\smallfigwidth]{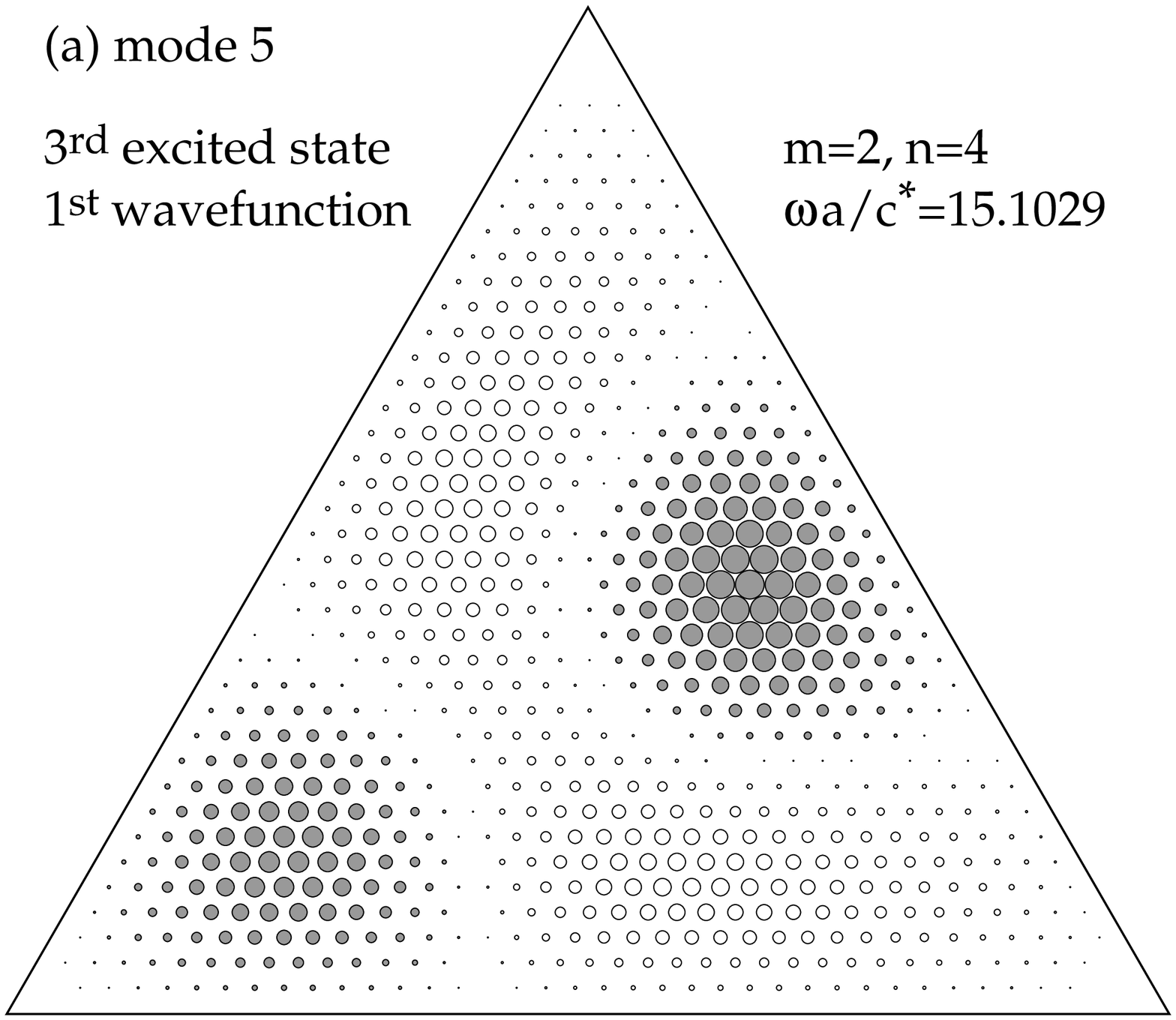}
\includegraphics[angle=0.0,width=\smallfigwidth]{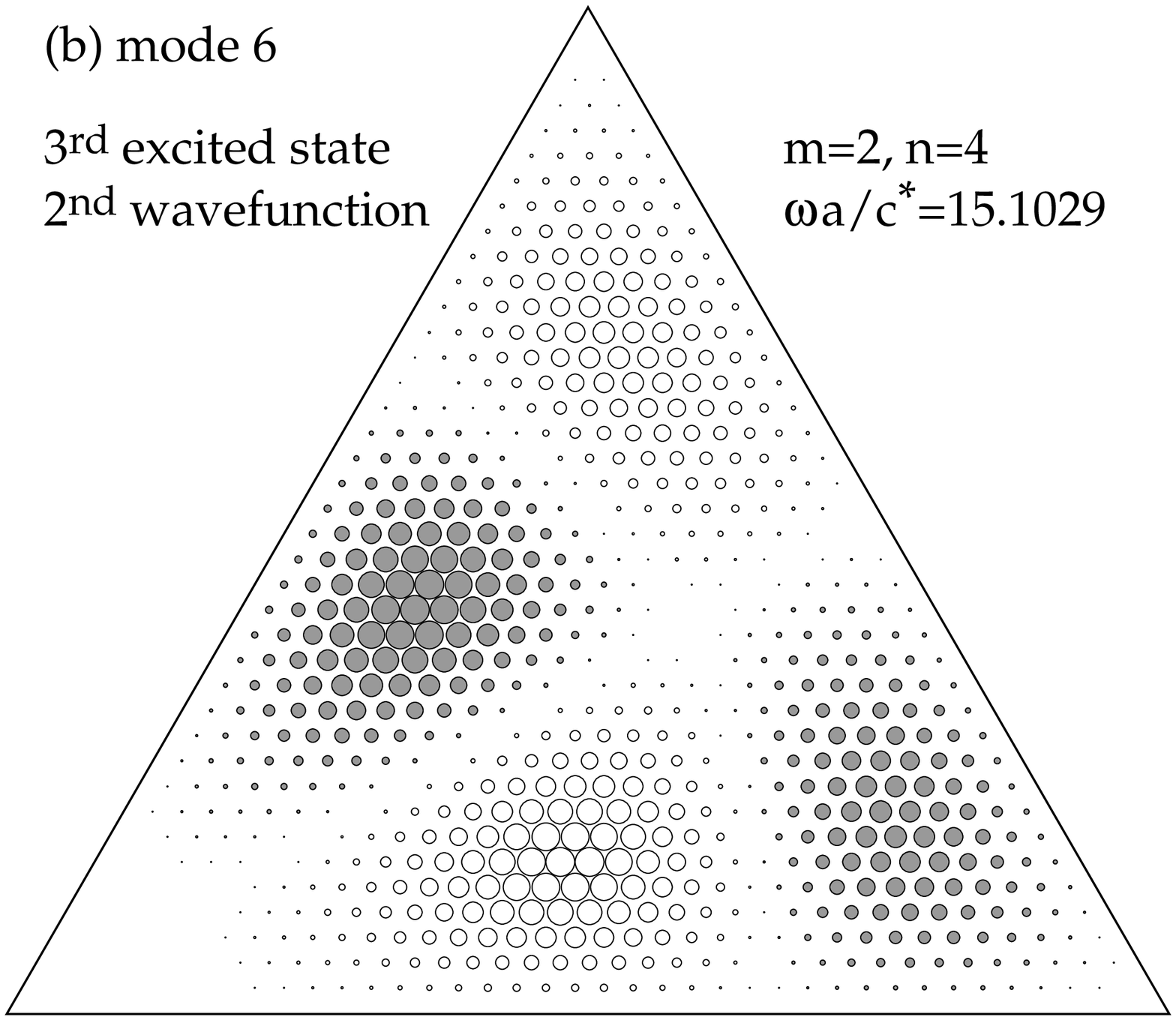}
\caption{
\label{trimodesC}
Wavefunctions of the doubly degenerate third excited state in a 2D 
triangular system.  Two orthogonal sub-states corresponding to 
different choices of the phase of $\psi$ are presented in (a) and (b).
}
\end{figure}

The wavefunctions of the lowest modes with $m=0$ are displayed in Fig.\ 
\ref{trimodesA}.  All the modes with $m=0$ have this typical
triangular shape, periodically patterned over the entire system.
Wavefunctions of the first two excited states are presented
in Figs.\ \ref{trimodesB} and \ref{trimodesC}.
Each of these states, having nonzero $m$, is doubly degenerate, therefore, 
there is some arbitrariness in the wavefunction pictures. 
Two degenerate wavefunctions were obtained from the real and imaginary
parts of $\psi$ in Eq.\ \eqref{triwf}; an equivalent degenerate pair
can be obtained by changing the complex constant ${\cal A}_0$.
The degenerate pairs can be shown to be orthogonal in the usual sense.
Wavefunctions for the sequence of many of the other lowest modes are displayed 
at \url{www.phys.ksu.edu/~wysin/}.
The modes described here are the complete set of allowed modes for the 
triangular geometry.

\subsection{Equilateral triangle with NBC}
Although we do not apply these in this work, it is interesting to note the minor 
differences in the solutions when the triangle is solved with Neumann BC.\cite{BB97}
As far as the changes in the wavefunctions, NBC simply requires the \emph{sine}
functions in the expressions for $\psi_0$, $\psi_1$, and $\psi_2$,
Eqs. \eqref{psi0}, \eqref{psi1}, \eqref{psi2}, to be replaced by \emph{cosine}
functions.
The expressions \eqref{k1values} and \eqref{k2values} for the allowed wavevectors
are still valid, except that the restriction $m<n$ should be changed to $m\le n$.
So the spectrum is expanded slightly, there is an entire set of nondegenerate
lower modes with $m=n$ (alternatively with rotated indexes $(m,0)$); the NBC ground 
state is lower than the DBC ground state.
%

\subsection{3D prisms}
For a 3D triangular-based prism of height $h$, and base edge $a$, 
the solutions to the wave equation can be written as products of 
an $xy$-dependent part $\psi_{xy}$ and a part $\psi_z$
depending only on the vertical coordinate $z$.
The corresponding $k$ and $k_z$ have already been discussed above. 
Solutions can be found combining  $k$ and $\psi_{xy}$ from the
2D solutions with the $k_z$ and $\psi_z$ discussed above.
This approach would assume DBC over the all surfaces of the cavity.
While the physical situation would not exactly satisfy the requirements for
classification of modes in TM and TE polarizations, the error in this assumption
should be least at small $k_z$. 
Thus, it could give some rough idea about the modification of the confined
modes due to the finite vertical height of the cavity.

\section{TIR confinement and the resonant mode spectrum}
Knowing the eigenmodes of the cavity, various questions can be addressed 
and answered based on the structure of the wavefunctions.
The primary question is:  which modes are confined by TIR (or resonant) when the 
previously assumed Dirichlet boundary conditions are replaced by an index mismatch 
at the boundary?
As seen below, some states can never be TIR confined.
States which have the possibility for confinement will be refered to as TIR states.
Related to this is to estimate a decay time for any of the TIR modes.
%

As suggested in the Introduction, the mode confinement can be determined
approximately by identifying those modes which completely satisfy
the elementary requirements for total internal reflection.
We assume that outside the system boundary a different refractive material with 
relative permeability $\mu'$ and permittivity $\epsilon'$ (which could be vacuum 
for greatest simplicity) and index $\mathsf{n}'=\sqrt{\epsilon' \mu'}$, rather 
than a conducting boundary.  
The wavefunction analysis above describes a mode as a superposition of plane
waves with the specified (full 3D) wavevectors $\vec{K}$. 
Each plane wave component can be analyzed to see whether it undergoes TIR
at all system boundaries on which it is incident.  
If all the plane wave components of a mode satisfy the TIR requirements, 
then the mode is confined; it should correspond to a resonance of the
optical cavity.  
If any of the plane wave components do not satisfy TIR, the fields
of the mode will quickly leak out of the cavity and there should be no
resonance at that mode's frequency (also, without TIR, the assumption of
DBC is completely invalidated).
These results clearly depend on the index ratio from inside to outside the cavity,
denoted as 
\begin{equation}
\mathsf{N}=\mathsf{\frac{n}{n'}}=\frac{\sqrt{\epsilon\mu}}{\sqrt{\epsilon'\mu'}}
\end{equation}
%

A plane wave within the cavity with (3D) wavevector $\vec{K}_i$ has an incident angle 
on one of the boundaries expressed as
\begin{equation}
\label{TIR1}
\sin\theta_i = \frac{K_{i\:||}}{K_i},
\end{equation}
where $||$ indicates the component parallel to the boundary.
TIR will take place for this wave provided 
\begin{equation}
\label{TIR}
\sin\theta_i > \sin\theta_c = \frac{1}{\mathsf{N}}.
\end{equation}
%

For 2D geometry,  Eqs.\ \eqref{TIR} is applied directly 
to 2D wavevectors, $\vec{K}_i=\vec{k}_i$: one just needs to identify the 
parallel component of $\vec{k}_i$ relative to a boundary, for each 
for each plane wave present in $\psi$, Eq.\ \eqref{triwf}. 
The main difficulty is to satisfy TIR on all boundaries on which that 
component is incident.  
%

For 3D prismatic geometry, the wavevector $\vec{K}_i$ includes a $z$-component.  
The net parallel component needed in \eqref{TIR}
can include planar ($xy$) and vertical ($z$) contributions.  
The addition of a $z$-component will have the tendency to raise the energy
of any mode, and the net wavevector magnitude, thereby improving the
possibility for mode confinement.  

\subsection{Confinement in 2D triangular systems} 
For both the TM and TE 2D modes of an equilateral triangle, we can use the exact 
solutions in Sec.\ \ref{Tri2D} to investigate whether these can be confined by TIR.
Due to the threefold rotational symmetry through angles of $0^{\circ}$,
$120^{\circ}$ and $-120^{\circ}$, the TIR condition need only be applied
on one of the boundaries, say, the lower boundary $b_0$, parallel to
the $x$-axis.
The first wave $\psi_0$ is composed of two traveling waves, one
incident on $b_0$ and one reflected from $b_0$; both have wavevectors 
whose $x$-component is $k_1=\frac{2\pi}{3a}m$, where $m=0,1,2,...$.
The other rotated waves $\psi_1$ and $\psi_2$ are composed from traveling
waves whose wavevectors have $x$-components of magnitudes 
$\frac{1}{2}(\sqrt{3}k_2 \pm k_1)$, where $k_2=\frac{2\pi}{3a}\sqrt{3}n$
with $n=1,2,3...$.
For the allowed modes, $m<n$, or $k_1<k_2$, which shows that
\begin{equation}
k_1 < \frac{1}{2}(\sqrt{3}k_2-k_1) \le \frac{1}{2}(\sqrt{3}k_2+k_1).
\end{equation}
The $x$-component of the $\psi_0$ wave is always the smallest; this wave has 
the smallest angle of incidence on $b_0$, so if it undergoes TIR then so do 
$\psi_1$ and $\psi_2$, and the mode is confined.
As $k_1$ is the component parallel to boundary $b_0$ (i.e., $k_{i\: ||}$),
the condition for TIR confinement of the mode is  
\begin{equation}
\label{triTIR1}
\sin\theta_i = \frac{k_1}{\sqrt{k_1^2+k_2^2}} > \sin\theta_c = \frac{1}{\sf N},
\end{equation}
or equivalently in terms of quantum numbers $m$ and $n$,
\begin{equation}
\label{triTIR2}
\frac{m}{n} > \sqrt{\frac{3}{{\sf N}^2 - 1}}.
\end{equation}
%

This relation contains some interesting features.
First, since all modes have $m<n$, the LHS is always less than $1$, and 
confinement of modes can only occur for adequately large refractive 
index ratio, $\mathsf{N}>2$.  
For $\mathsf{N}<2$ all modes will leak out of the cavity; they cannot
be stably maintained by TIR.
Secondly, for any particular value $\mathsf{N}>2$, relation \eqref{triTIR2}
determines a critical $m/n$ ratio;  modes whose $m/n$ ratio is below
the critical value will not be stable.   
As $m/n$ relates to the geometrical structure of the mode wavefunction, there 
is a strong relation between the confined mode structures and the refractive 
index.
An alternative way to look at this, is that each particular mode will be confined 
only if $\mathsf{N}$ is greater than a specific critical value determined by the 
$m/n$ ratio for that mode:
\begin{equation}
\label{triTIR_N}
\mathsf{N} > \mathsf{N}_c = \sqrt{3\frac{n^2}{m^2}+1}.
\end{equation}
The relation shows that generally speaking, modes with smaller $n/m$ are most
easily confined;  stated otherwise, the modes where $m$ is closest to $n$ are
the ones most readily confined.
On the other hand, all modes with $m=0$ leak out, because these have wave 
components with a vanishing angle of incidence on the boundaries.  

\begin{figure}
\includegraphics[angle=-90.0,width=\columnwidth]{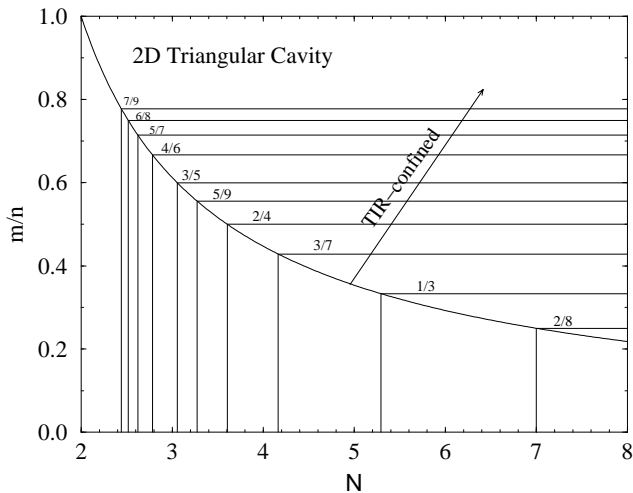}
\caption{
\label{trifig1}
TIR mode confinement limits (index ratio 
$\mathsf{N}=\sqrt{\epsilon\mu/\epsilon' \mu'}$) for 2D triangular cavities.
Modes are confined where the $m/n$ ratios (as indicated) lie above the 
solid curve, corresponding to TIR on all boundaries [relation \eqref{triTIR2}].  
Intersections on the $\mathsf{N}$-axis give the critical index ratios for each mode.
}
\end{figure}

Fig.\ \ref{trifig1} is a kind of phase diagram for mode confinement by TIR.
Some of the modes' $m/n$ ratios are plotted vs. refractive index ratio
$\mathsf{N}$, together with relation \eqref{triTIR2}.
A particular mode is confined only if its $m/n$ value falls above the critical
curve.  
In this way one can easily see the critical refractive indexes for each mode.
As mentioned above, modes where $m$ is close to $n$ require the smallest
refractive index values for confinement.  

\begin{figure}
\includegraphics[angle=-90.0,width=\columnwidth]{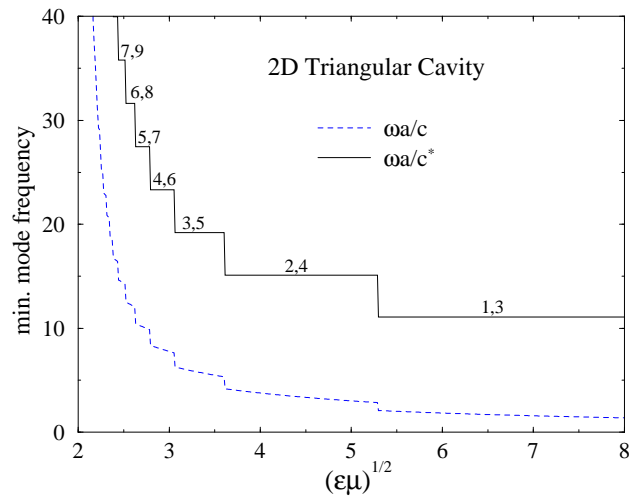}
\caption{
\label{trifig2}
Frequency of the lowest confined mode for a 2D triangular system surrounded
by vacuum, as a function of the refractive index.  Pairs $(m,n)$ indicate some 
of the modes' quantum numbers. No modes are confined for 
$\sqrt{\epsilon\mu}\le 2$.  
}
\end{figure}

One more application of these results is shown in Fig.\ \ref{trifig2},
where the frequency of the lowest confined mode is plotted versus 
$\sqrt{\epsilon\mu}$, with vacuum outside the cavity ($\epsilon' \mu'=1$).
The quantum numbers $(m,n)$ for the lowest confined mode are indicated in 
each curve segment.
The two different curves correspond to plotting $\omega a/c$ (dashed)
and frequency scaled with refractive index, $(\omega a/c)\sqrt{\epsilon\mu}$
(solid).
Of course, as $\sqrt{\epsilon\mu}$ approaches the value $2$, the lowest 
frequency becomes large and the minimizing mode has $m$ very close to $n$, 
with both large. 
At the opposite limit of large values of $\sqrt{\epsilon\mu}$, the lowest 
frequency becomes small, and the mode $(1,3)$ is always the lowest frequency
mode that is confined for any $\sqrt{\epsilon\mu}>5.29$ . 
The graph only shows the minimum confined frequency at the intermediate 
values of refractive index $\sqrt{\epsilon\mu}$.

\subsection{Confinement in triangular based prisms}
We consider a vertical prism of height $h$ with a triangular base of
edge $a$.
With perfectly reflecting end mirrors at $z=0$ and $z=h$ the allowed longitudinal
wavevectors would be $k_z = l\pi/h$, where $l$ is an integer.
For a plane wave component with 2D wavevector $\vec{k}_i$, the net 3D wavevector 
magnitude within the cavity is
\begin{equation}
K_i = \sqrt{k_i^2+k_z^2}.
\end{equation}
This increases the incident angles on the walls of the cavity: the presence of 
nonzero $k_z$ should enhance the possibility for confinement, compared to the 
purely 2D system.

At the lower and upper ends of the prism, the parallel wavevector
component of any of the plane waves present is the full 2D $\vec{k}_i$, 
which can be expressed as
\begin{equation}
\label{t-ends}
K_{i\:||}^{\textrm{ (ends)}}=|\vec{k}_i|=\sqrt{k_1^2+k_2^2}
=\frac{2\pi}{3a}\sqrt{m^2+3n^2}.
\end{equation}
Finding the incident angle by \eqref{TIR1} and Snell's Law \eqref{TIR},
this implies a critical index ratio needed for TIR confinement by the cavity ends,
\begin{equation}
\label{tri-ends}
\mathsf{N}_c^{\textrm{ (ends)}} = \frac{1}{\sin\theta_i}=
\sqrt{1+\left(\frac{3}{2\pi}\right)^2 \frac{(k_z a)^2}{m^2+3n^2}}.
\end{equation}
%

On the vertical walls of the prism, the parallel wavevector component
is a combination of $k_z$ and the 2D $k_{i\:||}$.
If TIR occurs on on one wall then by symmetry it will take place
on all the walls.
Considering the $b_0$ wall, the $\psi_0$ wave has the largest incident angle
as in the 2D problem, and both $k_1$ and $k_z$ determine its parallel 
wavevector component,
\begin{equation}
\label{t-walls}
K_{i\:||}^{\textrm{ (walls)}}=
\sqrt{k_1^2+k_z^2}=\sqrt{\left(\frac{2\pi}{3a}\right)^2 m^2+k_z^2}.
\end{equation}
Then this determines the critical index ratio needed for TIR by the
cavity walls,
\begin{equation}
\label{tri-walls}
\mathsf{N}_c^{\textrm{ (walls)}} = \frac{1}{\sin\theta_i}= 
\sqrt{1+\frac{3n^2}{m^2 + (\frac{3}{2\pi})^2 (k_z a)^2} }.
\end{equation}
For TIR mode confinement, the actual index ratio $\mathsf{N}$ must be
greater than both $\mathsf{N}_c^{\textrm{ (ends)}}$ and 
$\mathsf{N}_c^{\textrm{ (walls)}}$.

\begin{figure}
\includegraphics[angle=-90.0,width=\columnwidth]{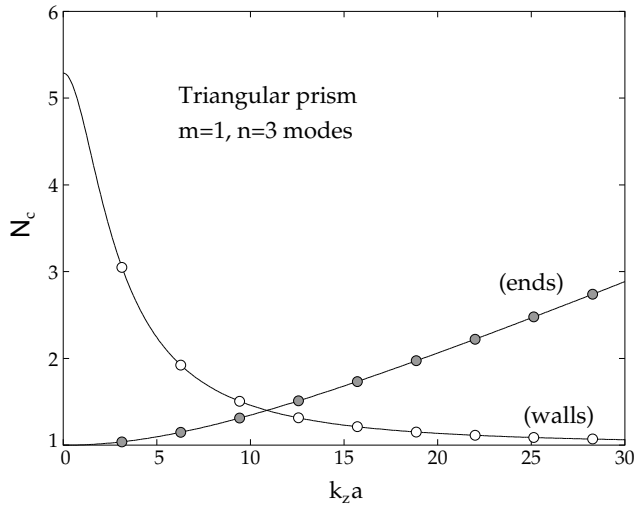}
\caption{
\label{triprism13}
Triangular prism critical index ratios for TIR from the walls (open symbols)
and ends (solid symbols) of the cavity, for the modes with 2D quantum indexes
$m=1$, $n=3$, and longitudinal wavevector $k_z$.  The symbols show the 
allowed $k_z a$ values $l\pi a/h$ when $h=a$.  
}
\end{figure}

\begin{figure}
\includegraphics[angle=-90.0,width=\columnwidth]{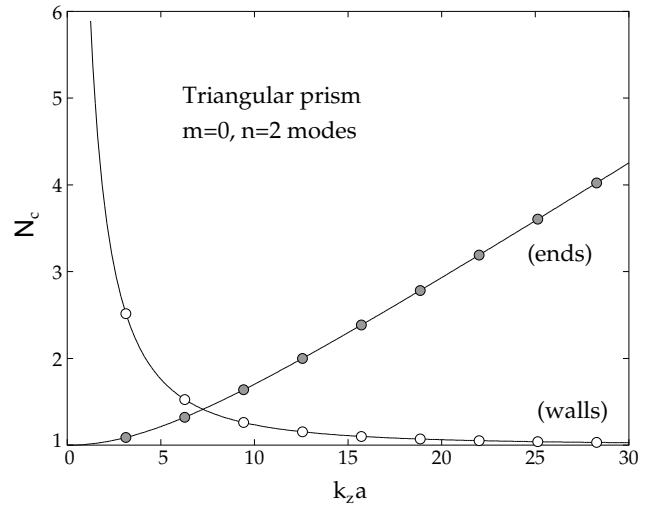}
\caption{
\label{triprism02}
Triangular prism critical index ratios as explained in Fig.\ \protect\ref{triprism13}
for the modes with 2D quantum indexes $m=0$, $n=2$ (2D ground state).
}
\end{figure}

Some results are shown in Fig.\ \ref{triprism13} for the $m=1$, $n=3$ 
modes as a function of $k_z$.
The symbols indicate the allowed $k_z$ values for the prism
height equal to the base edge, $h=a$. 
The limiting 2D critical ratio is seen at the point $k_z a=0$.
The critical index ratio is lowered from the 2D value $\mathsf{N}_c\approx 5.29$,
down to values as low as $\mathsf{N}_c\approx 1.40$ for $k_z a\approx 11$. 
A more significant reduction in $\mathsf{N}_c$ occurs for the ground state, 
$m=0$, $n=2$, Fig.\ \ref{triprism02}.
It is not possible to confine this mode in 2D; whereas, it has
$\mathsf{N}_c \approx 1.41$ for $k_z a\approx 7.25$.  
Similar significant critical index ratio reductions occur for all the modes 
tested.

One sees that $\mathsf{N}_c^{\textrm{ (walls)}}$ and $\mathsf{N}_c^{\textrm{ (ends)}}$
always cross at some intermediate $k_z$, where they produce the smallest
index ratio $\mathsf{N}_c^{\textrm{ (min)}}$ needed for TIR. 
Setting them equal, we find
\begin{equation}
\label{Nmin}
\mathsf{N}_c^{\textrm{ (min)}} = \sqrt{1+\left(1+\frac{m^2}{3n^2}\right)^{-1}},
\end{equation}
which occurs at 
\begin{equation}
k_z^{*} = \left(\frac{2\pi}{3a}\right)\sqrt{3}n = k_2.  
\end{equation}
It is a rather intriguing result; choosing $h/a$ such that a mode will 
occur at this $k_z$ will be the optimum choice for having the mode confined
most easily by TIR.
This could be useful for control over selection of desired modes in a cavity.

Cautionary comments are in order.
The above discussion applies exactly only to a scalar wave.
For electromagnetic waves, it applies to either the TM or TE polarizations in an 
approximate sense for quasi-2D EM modes, requiring small longitudinal 
wavevector $k_z a<1$. 
The greatest enhancements in TIR were found to occur at values $k_z a>1$,
primarily because a reasonably large $k_z$ is needed in order to satisfy
the TIR requirements at the ends.  
The presence of significant $k_z$ values, however, will lead to a mixing
of the TM and TE polarizations, making this calculation invalid.
Then, for practical purposes, the calculation is mostly interesting for
how it indicates the improvement in TIR confinement expected mainly 
on the 2D walls of the cavity, at small $k_z$.
%

\section{TIR mode lifetimes}
For the 2D solutions ($k_z=0$),  it is interesting to estimate the lifetimes
of the TIR-confined modes, contrasting the results for TM and TE polarizations.
When all of the plane wave components in $\psi$ satisfy the TIR conditions,
there is still the possibility for the cavity fields to decay in time.
Clearly, we have only an approximate solution, since DBC is not exactly
the correct boundary condition.  
The effect this causes is difficult to estimate.  
Another source of decay are diffractive effects: the finite length of the
triangle edge and the presence of sharp corners is likely to have special
influence on the TIR that is difficult to predict.
One feature, however, which can be considered as due to diffraction, 
is the leakage of \emph{boundary waves} at the corners of the 
triangle.\cite{Wiersig03}
Under conditions of TIR, an evanescent wave exists within the exterior
medium, decaying exponentially into that medium, and moving parallel
to the cavity surface.
When it encounters the corner of that edge, a sharp discontinuity in the surface,
it can be expected to constitute power radiated from the cavity. 
Here we consider the mode lifetime estimates based solely on the losses due to
these boundary waves.

Based on the ratio of the total energy $U$ stored in the cavity fields,
compared to the total power $P$ emitted by the boundary waves from all the corners, 
an upper limit of the mode lifetime can be estimated as
\begin{equation}
\label{taudef}
\tau = \frac{U}{P}.
\end{equation}
The calculations of $U$ and $P$ have slight differences for TM versus TE polarization.
Therefore, there is no reason to expect these lifetimes to be the same.

\textbf{Cavity energy:}
For both polarizations, we use the wavefunction $\psi$ reviewed in Sec.\ \ref{Tri2D},
which can be expressed more succinctly as
\begin{eqnarray}
\label{psi}
\psi &=& 
{\cal A}_0 \left\{ e^{i k_1 x} 
           \sin\Big[k_2\big(y+\frac{a}{2\sqrt{3}}\big)\Big] \right.  \\
 &+& e^{ik_1(-\frac{1}{2}x+\frac{\sqrt{3}}{2}y+a)}
           \sin\Big[k_2\big(-\frac{\sqrt{3}}{2}x-\frac{1}{2}y+\frac{a}{2\sqrt{3}}\big)\Big] 
   \nonumber  \\
 &+& \left.
    e^{ik_1(-\frac{1}{2}x-\frac{\sqrt{3}}{2}y-a)}
           \sin\Big[k_2\big(\frac{\sqrt{3}}{2}x-\frac{1}{2}y+\frac{a}{2\sqrt{3}}\big)\Big] 
 \right\} \nonumber 
\end{eqnarray}
The total energy of the fields within the cavity of height $h$ can be written as
\begin{equation}
\label{U}
U=\int h~ dx dy ~ \frac{\epsilon |\vec{E}|^2}{8\pi}  
 =\int h~ dx dy ~ \frac{|\vec{B}|^2}{8\pi\mu};
\end{equation}
the first form is convenient for TM modes ($|\vec{E}|^2=|\psi|^2$),
the second is convenient for TE modes ($|\vec{B}|^2=|\psi|^2$).
So both calculations require the normalization integral of $\psi$.
This integral can be simplified by a transformation to a skew coordinate system whose 
axes are aligned to two edges of the triangle as shown in Fig.\ \ref{trisys}.
Placing the origin of the new coordinates $(X,Y)$ at the lower left corner of the
triangle, with $X$ increasing from $0$ to $a$ along edge $b_0$, and $Y$ increasing from
$0$ to $a$ along edge $b_2$, the definition of the new coordinates results from
\begin{eqnarray}
X + Y\cos 60^{\circ} &=& x+\frac{a}{2}, \nonumber \\
Y \sin 60^{\circ} &=& y+\frac{a}{2\sqrt{3}}.
\end{eqnarray}
The numbers $(\frac{a}{2}, \frac{a}{2\sqrt{3}})$ are simply the displacement
of the origin (vector from triangle corner $L$ to center $C$).
The wavefunction is now expressed as
\begin{eqnarray}
\label{psiXY}
\psi &=& {\cal A}_0 \left\{ e^{ik_1 \big(X+\frac{1}{2}Y-\frac{a}{2}\big)}
                            \sin\Big[\frac{\sqrt{3}}{2}k_2 Y\Big] \right. \nonumber \\
     &+&  e^{ik_1 \big(-\frac{1}{2}X+\frac{1}{2}Y+a\big)}
           \sin\Big[\frac{\sqrt{3}}{2}k_2 \big(-X-Y+a)\Big] \nonumber \\
     &+&  \left. e^{ik_1 \big(-\frac{1}{2}X-Y-\frac{a}{2}\big)}
           \sin\Big[\frac{\sqrt{3}}{2}k_2 X \Big] \right\}.
\end{eqnarray}
In this form, it is more obvious that each term goes to zero on one of
the boundaries, $X=0$ ($b_2$), $Y=0$ ($b_0$), or $X+Y=a$ ($b_1$). 
Using the periodicity of $\psi$, the integration over the triangular area is effected by 
\begin{equation} 
\int dx \; dy = \frac{1}{2} \int_{0}^{a} dX \int_{0}^{a} dY ~ |J|
\end{equation}
where the Jacobian is $|J|=\frac{\sqrt{3}}{2}$ and the factor of $\frac{1}{2}$
cancels integrating over two triangles.
The absolute square of $\psi$ involves three direct terms (squared sines involving 
only $k_2$) and six cross terms from Eq.\ \eqref{psiXY}.
It is possible to show that the cross terms integrated over the triangular area
all are zero, due to the special choices of allowed $k_1$ and $k_2$ given by 
\eqref{k1values} and \eqref{k2values}.
The remaining nonzero parts result in
\begin{equation}
\label{psinorm}
\int dx \; dy ~ |\psi|^2 =  \frac{3\sqrt{3}}{8} a^2 |{\cal A}_0|^2  .
\end{equation}
%

{\bf Boundary wave power:}
The symmetry of the wavefunction causes the boundary wave power out of each
edge to be the same, therefore, we calculate that occurring in edge $b_0$ 
(at $y=0$) and multiply by three for the total power.  This calculation
follows that presented by Wiersig\cite{Wiersig03} for resonant fields in a
regular polygon.

Looking at the wavefunction \eqref{psi}, one can see that there are three distinct
plane waves \emph{incident} on $b_0$.
First is the wave with the smallest angle of incidence, resulting from the first term
in \eqref{psi},
\begin{equation}
\label{psi0-}
\psi_{0}^{-} = \frac{-{\cal A}_0}{2i} e^{\frac{-ik_2 a}{2\sqrt{3}}} e^{i(k_1 x-k_2 y)}.
\end{equation}
Using the allowed values for $k_1$ and $k_2$, the angle of incidence is seen to be
\begin{equation}
\label{thetaTM}
\sin\theta_{0}^{-} = \frac{m}{\sqrt{m^2+3 n^2}}.
\end{equation}
Next, there is a wave with the largest magnitude incident angle, due to the second
term in \eqref{psi},
\begin{equation}
\label{psi1+}
\psi_{1}^{+} = \frac{{\cal A}_0}{2i} e^{i(k_1+\frac{k_2}{2\sqrt{3}})a} 
 e^{i[(-\frac{1}{2}k_1-\frac{\sqrt{3}}{2}k_2)x+(\frac{\sqrt{3}}{2}k_1-\frac{1}{2}k_2)y]},
\end{equation}
whose incident angle is
\begin{equation}
\sin\theta_{1}^{+}= \frac{1}{2} \frac{-m-3n}{\sqrt{m^2+3n^2}}.
\end{equation}
A negative value of $\theta_{1+}$ means the wave is propagating contrary to the
$x-axis$.
Finally, the last term in \eqref{psi} leads to a wave with an intermediate
incident angle,
\begin{equation}
\label{psi2+}
\psi_{2}^{+} = \frac{{\cal A}_0}{2i}e^{i(-k_1+\frac{k_2}{2\sqrt{3}})a}
 e^{i[(-\frac{1}{2}k_1+\frac{\sqrt{3}}{2}k_2)x+(-\frac{\sqrt{3}}{2}k_1-\frac{1}{2}k_2)y]},
\end{equation}
whose incident angle is
\begin{equation}
\sin\theta_{2}^{+}= \frac{1}{2} \frac{-m+3n}{\sqrt{m^2+3n^2}}.
\end{equation}
The plus/minus superscripts on these waves refer to the positive/negative exponents in the
sine functions of Eq.\ \eqref{psi}. 

Now, for each of these incident waves, there is a corresponding evanescent wave
propagating along the edge of the cavity; these are assumed to produce emitted power
when encountering the triangle corners.
The Poynting vector $\vec{S}'$ associated with a single plane evanescent wave along 
the $b_0$ boundary is 
\begin{equation}
\label{Poynting}
\vec{S}' = \frac{c}{8\pi} \Re (\vec{E}'\times\vec{H}'^{*})
         = \frac{c}{8\pi} \sqrt{\frac{\epsilon'}{\mu'}}|\vec{E}'|^2 \sin\theta' ~ \hat{x}
\end{equation}
On the other hand, the linear superposition of the three waves 
$\psi_{0}^{-}, \psi_{1}^{+}, \psi_{2}^{+}$ leads to an interference pattern both within the 
cavity, and in the evanescent waves and exterior power flow.
Careful consideration of a linear combination of two waves shows that, although 
interference leads to a spatially varying $\vec{S}'$ with components both parallel 
and perpendicular to the boundary, an integral number of wavelengths of that 
pattern fits along the edge.
Thus, it is clear that the interference effects can be ignored in the calculation
of the emitted boundary power.
For the total emitted power due to boundary waves, it is sufficient to sum the
individual powers for the three independent incident waves.
 
For \textbf{TM polarization}, with $\psi=E_z$, the exterior electric field of a 
single evanescent wave has only a $z$-component like that in Eq.\ \eqref{E'TM}.
Applying Snell's law, and integrating the Poynting vector from $y=0$ to $y=\infty$, 
the power flow along $\hat{x}$ in one boundary wave, on one edge, is
\begin{equation}
\label{TMBWP}
P_x = \frac{ch}{4\pi\mu'}
\frac{|E_i^{0}|^2}{\omega/c}
\frac{\mathsf{n}\sin\theta_i}{\sqrt{(\mathsf{n}\sin\theta_i)^2-(\mathsf{n}')^2}} 
\cos^2 \frac{\alpha}{2}
\end{equation}
where $\alpha$ is the phase shift given by \eqref{alphaTM} in Sec.\ \ref{2DEM}.
One can see that the boundary wave power has a dependence on
$(\sin\theta_i-\sin\theta_c)^{-1/2}$.
Although each of the waves $\psi_{0}^{-}, \psi_{1}^{+}, \psi_{2}^{+}$ will produce a
boundary wave emission, the wave $\psi_{0}^{-}$ has the smallest angle of incidence,
and produces by far the largest boundary wave power.
This can be seen by examining the expressions for incident angles 
$\theta_{0}^{-}, \theta_{1}^{+}$ and $\theta_{2}^{+}$, and using the facts that 
$m<n$ and $m,n$ have the same parity.
Therefore a good lifetime estimate can be made using only the power due to $\psi_{0}^{-}$.
From the expression \eqref{psi0-} for $\psi_{0}^{-}$,  the squared magnitude of
its electric field is $|E_i^{0}|^2 = \frac{1}{4}|{\cal A}_0|^2$. 
From expressions \eqref{psinorm} and \eqref{U}, the total TM cavity energy is
\begin{equation}
U_{\textrm{TM}} = \frac{\epsilon h}{8\pi} \frac{3\sqrt{3}}{8} a^2 |{\cal A}_0|^2 .
\end{equation}
The estimate of the lifetime due to boundary wave emission only, from
all three edges combined,  is $\tau_{\textrm{TM}} \approx U_{\textrm{TM}}/3 P_x$.
It is convenient to express the reult in dimensionless form, scaling with the mode frequency,
\begin{eqnarray}
\label{tauTM}
\omega\tau_{\textrm{TM}} &\approx&  \frac{\sqrt{3}}{4} \left(\frac{\omega a}{c^*} \right)^2
 \frac{\sqrt{1 - (\sin\theta_c/\sin\theta_{0}^{-})^2}}
      {\cos^2 \theta_{0}^{-}} 
 \nonumber \\
 &\times& \frac{\mu}{\mu'} 
    \left[ \sin^2\theta_{0}^{-} - \sin^2\theta_c + 
           \left(\frac{\mu'}{\mu}\right)^2 \cos^2\theta_{0}^{-} \right],
\end{eqnarray}
where $\theta_{0}^{-}$ depends on the mode quantum numbers according to 
Eq.\ \eqref{thetaTM}.
In the usual case where $\mu\approx\mu'$, the second line of the formula
simplifies to just $\cos^2 \theta_c$.

Obviously, when $\theta_{0}^{-}$ approaches $\theta_c$, which would occur at
weak enough index mismatch, the estimated lifetime $\tau_{\textrm{TM}}\to 0$, 
which is the limit of a non-bound state.   
At the opposite extreme of large index mismatch where $\mathsf{n/n}' \gg 1$, 
this estimate varies as $1/\cos^2 \theta_{0}^{-}$, and since $\omega a/c^*$ 
is a number of order unity, the order of magnitude is determined by
\begin{equation}
\label{tau_sim}
\tau_{\textrm{TM}} \sim \frac{a}{c} \sqrt{\epsilon\mu} .
\end{equation}
The result is interesting because it shows a lifetime that increases with the 
triangle size, as well as being proportional to the refractive index in the cavity.

For \textbf{TE polarization}, with $\psi=B_z$, the exterior magnetic field of a 
single evanescent wave has only a $z$-component like that in Eq.\ \eqref{B'TE}.
The calculation follows the same reasoning as used for the TM modes, but
the specific details lead to a slightly different result. 
Definition of the Poynting vector as in \eqref{Poynting} is the same.
The substitution $|\vec{E}'|=|B_z'|/\sqrt{\epsilon'\mu'}$ together with
Eq.\ \eqref{B'TE} for $B_z'$, followed by $|E_i^{0}|=|B_i^{0}|/\sqrt{\epsilon\mu}$
within the cavity, leads to the boundary wave power with extra factors,
\begin{equation}
\label{TEBWP}
P_x = \frac{ch}{4\pi\mu'} \frac{\mu'\epsilon}{\epsilon'\mu}
\frac{|E_i^{0}|^2}{\omega/c}
\frac{\mathsf{n}\sin\theta_i}{\sqrt{(\mathsf{n}\sin\theta_i)^2-(\mathsf{n}')^2}} 
\cos^2 \frac{\alpha}{2}
\end{equation}
where now the phase shift $\alpha$ given by \eqref{alphaTE} in Sec.\ \ref{2DEM}
depends on the ratio $\epsilon/\epsilon' \gg 1$ rather than $\mu/\mu'\approx 1$.
Furthermore, using \eqref{psinorm} and \eqref{U}, the energy stored in the 
cavity fields is now
\begin{equation}
U_{\textrm{TE}} = \frac{h}{8\pi\mu} \frac{3\sqrt{3}}{8} a^2 |{\cal A}_0|^2 .
\end{equation}
With cavity magnetic field strength $|B_i^{0}|^2 = \frac{1}{4}|{\cal A}_0|^2$,
and again estimating the lifetime using only the $\psi_{0}^{-}$ boundary wave
from all three edges, the lifetime is 
$\tau_{\textrm{TE}} \approx U_{\textrm{TE}}/3 P_x$.
One finds
\begin{eqnarray}
\label{tauTE}
\omega\tau_{\textrm{TE}} &\approx&  \frac{\sqrt{3}}{4} 
\left(\frac{\omega a}{c^*} \right)^2
 \frac{\sqrt{1 - (\sin\theta_c/\sin\theta_{0}^{-})^2}}
      {\cos^2 \theta_{0}^{-}}
 \nonumber \\
 &\times& \frac{\epsilon}{\epsilon'} 
    \left[ \sin^2\theta_{0}^{-} - \sin^2\theta_c + 
           \left(\frac{\epsilon'}{\epsilon}\right)^2 \cos^2\theta_{0}^{-} \right].
\end{eqnarray}
The second line of this formula highlights the difference for TE polarization
compared to TM.   
The factor in the brackets is some number less than 1; it contrasts the
bracket which reduces to $\cos^2\theta_c$ in formula \eqref{tauTM} for the 
TM polarization lifetime when $\mu=\mu'$.
The crucial difference is the factor $\epsilon/\epsilon' \gg 1$
present here, compared to a similar factor $\mu'/\mu \approx 1 $ for the TM
lifetime formula.
This is the more dominant factor, and it suggests that roughly speaking, the
ratio of the lifetimes for the two polarizations, which have the same (approximately 
DBC) boundary conditions and frequencies,  is
\begin{equation}
\label{tauratio}
\frac{\tau_{\textrm{TE}}}{\tau_{\textrm{TM}}} \approx \frac{\epsilon}{\epsilon'} 
   = \left(\mathsf{\frac{n}{n'}}\right)^2 .
\end{equation}
The result holds as long as the index mismatch is adequately large compared 
to the cutoff value needed to stabilize that mode by TIR.  
Otherwise, at smaller index mismatch, the TM lifetime can be longer than
the TE lifetime.

\begin{figure}
\includegraphics[angle=-90.0,width=\columnwidth]{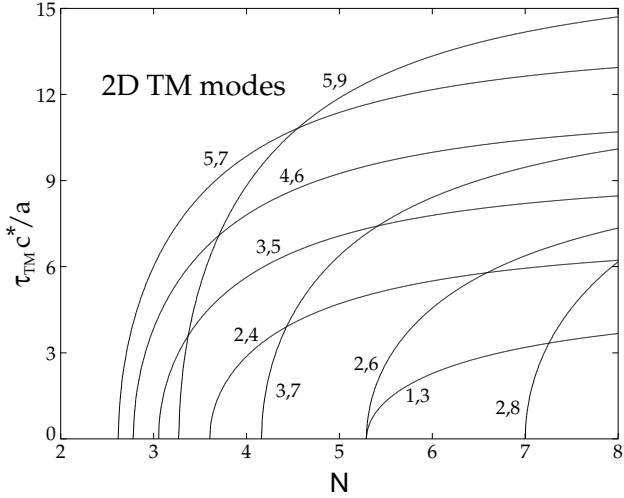}
\caption{
\label{tauTMfig}
Estimated lifetimes for some low TM modes indicated by $(m,n)$ pairs, 
versus the index ratio $\mathsf{N}$.  The lifetime is scaled by the speed 
of light $c^*$ in the cavity medium and the cavity edge size $a$.
}
\end{figure}

Some results for $\tau_{\textrm{TM}}$ are presented in Fig.\ \ref{tauTMfig},
showing lifetimes as functions of the index mismatch for some of the lowest modes.
When scaled by the triangle size and light speed in the cavity, the lifetimes
increase abruptly above the TIR confinement limits, eventually increasing at
a slower rate.  
For the modes shown, dimensionless frequencies $\omega a/c^*$ are typically numbers 
greater than 10 (see Fig.\ \ref{trifreqs}), with the values of $\tau c^*/a$ also of
the order of 10.
Combining these rough results, the mode lifetimes in units of the mode periods $T$ are
similar to
\begin{equation}
\frac{\tau}{T}=\frac{\omega\tau}{2\pi} \sim \frac{10 \times 10}{2\pi} \approx 16.
\end{equation}
Assuming that the lifetime estimates have included the dominant loss mechanism
in the cavity, this large result for $\tau/T$ indicates that the original assumption
of a resonance mode weakly confined by TIR should be a valid concept.
Obviously, this holds far enough above the TIR confinement limits only, keeping
in mind the approximate nature of the Dirichlet boundary conditions that were applied.  

\begin{figure}
\includegraphics[angle=-90.0,width=\columnwidth]{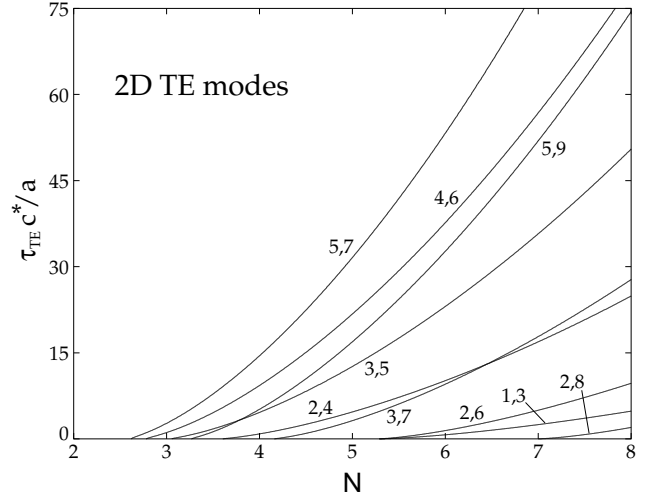}
\caption{
\label{tauTEfig}
Estimated lifetimes for low TE modes indicated by $(m,n)$ pairs, 
versus the index ratio $\mathsf{N}$.  These lifetimes are larger 
than the corresponding TM lifetimes (Fig.\ \protect\ref{tauTMfig})
when $\mathsf{N}$ is sufficiently larger than the cutoff value for that mode.
}
\end{figure}

Comparitive results for $\tau_{\textrm{TE}}$ for the same mode indexes are shown
in Fig.\ \ref{tauTEfig}.
Due to the presence of the extra factor of $\epsilon/\epsilon'\approx \mathsf{N}^2$, 
these lifetimes increase much more rapidly than the TM lifetimes.
As an example, for mode $(3,5)$, the TE lifetime is about 6 times
longer than the TM lifetime at $\mathsf{N}=8$.
On the other hand, for the mode $(1,3)$, which has a much larger TIR confinement
limit, the TE lifetime is only about 1/3 longer than the TM lifetime at 
$\mathsf{N}=8$.
Clearly, modes with $(m,n)$ indexes nearly the same, as in the form $(m,m+2)$
with large $m$, require smaller index mismatch for TIR confinement, and 
will more closely follow the lifetime ratio determined strongly by
the index mismatch, Eq.\ \eqref{tauratio}.

In recent experiments\cite{Chang00} triangular semiconductor cavities with
edges ranging from 75 to 350 $\mu$m were used.   
Assuming an effective index of refraction around $\mathsf{n}\approx 4$, with
vacuum on the exterior, and using $a\approx 100 \mu$m, Eqs.\ \eqref{tau_sim}
and \eqref{tauratio} give rough lower estimates $\tau_{\textrm{TM}}\sim 1.3$ ps,
and $\tau_{\textrm{TE}} \sim 20$ ps.  
Of course, for practical purposes of maintaining a resonating mode,
a large value of $\tau/T$ is much more relevant, as discussed above.
Based on these calculations, smaller cavities will have reduced lifetimes,
but also shorter oscillation periods in the same ratio.
For large index ratio, where $\sin\theta_c \ll 1$, $\cos\theta_c\approx 1$, and using 
the fact that $m<n$, lifetime expression \eqref{tauTM} and frequency expression
\eqref{2domega} produce the estimate, 
\begin{equation}
\frac{\tau_{\textrm{TM}}}{T} \approx \frac{\pi^2 n^2}{\sqrt{3}}.
\end{equation}
This ratio is independent of the cavity size or dielectric properties, increasing
only with the squared mode quantum number $n$.
The TE mode lifetime under these assumptions should be even larger, by the squared
refractive index ratio $(\mathsf{n/n'})^2$.

\section{Conclusions}
The resonant modes of an optical cavity with an equilateral 2D cross-section
have been examined in an approximate manner, starting from the exact analytic
solutions for Dirichlet boundary conditions.
The six plane waves that make up each mode have been analyzed for the conditions
necessary such that all are confined in the cavity by TIR.
For 2D electromagnetics ($k_z=0$), modes with larger ratios of quantum indexes 
$m/n$ are most easily confined; conversely, modes with $m=0$ never undergo TIR 
confinement.
The presence of a nonzero longitudinal wavevector $k_z$ can be expected to
improve the possibilities for TIR confinement.

The critical index ratios for TIR confinement are the same for TM and TE
polarizations.
The different Fresnel factors, however, produce different rates of energy lost
from the cavity by the evanescent boundary waves.
For the 2D problem, this was shown to lead to considerably longer lifetimes
of the TE polarization, enhanced approximately by a factor of the squared index
ratio $(\mathsf{n/n'})^2$ conpared to the TM lifetime.
The differences in these lifetimes would be expected to imply stronger coupling 
of EM fields from outside to inside the cavity in the TM polarization; it
suggests that stimulation and generation of the modes by an external light source 
should be more efficient for TM polarization.

\bibliography{waveoptics}

\end{document}